\documentclass[%
 aip,
 amsmath,amssymb,
preprint,%
]{revtex4-1}

\usepackage{graphicx}
\usepackage{dcolumn}
\usepackage{bm}
\usepackage{siunitx}
\usepackage[utf8]{inputenc}
\usepackage[T1]{fontenc}
\usepackage{mathptmx}
\usepackage{etoolbox}

\makeatletter
\def\@email#1#2{%
 \endgroup
 \patchcmd{\titleblock@produce}
  {\frontmatter@RRAPformat}
  {\frontmatter@RRAPformat{\produce@RRAP{*#1\href{mailto:#2}{#2}}}\frontmatter@RRAPformat}
  {}{}
}%
\makeatother

\usepackage{tikz}
\usepackage{epsfig}
\usepackage{epstopdf}
\usepackage{siunitx}
\usepackage{pbox}
\usepackage{makecell}
\usepackage{multirow}
\usepackage{amssymb}
\usepackage{amsmath}
\usepackage[normalem]{ulem}

\usepackage{blindtext}
\usepackage{etoolbox}
\usepackage[bottom]{footmisc}
\usepackage{fancyhdr}
 
\begin{document}

%

\title{Schottky-like barrier characterization of field-effect transistors with multiple quasi-ballistic channels}

\author{Anibal Pacheco-Sanchez*}%
\email{AnibalUriel.Pacheco@uab.cat.}
\affiliation{ 
Departament d’Enginyeria Electrònica, Escola d’Enginyeria, Universitat Autònoma de Barcelona, Campus UAB, 08193 Bellaterra, Spain
}%
\author{Quim Torrent}%
\affiliation{ 
Facultat de Ciències, Universitat Aut\`{o}noma de Barcelona, 08193 Bellaterra, Spain
}%
\author{David Jiménez}
\affiliation{ 
Departament d’Enginyeria Electrònica, Escola d’Enginyeria, Universitat Autònoma de Barcelona, Campus UAB, 08193 Bellaterra, Spain
}


%

\begin{abstract}
The potential barrier height at the interface formed by a metal contact and multiple one-dimensional (1D) quasi-ballistic channels in field-effect transistors (FETs) is evaluated across different carbon nanotube and nanowire  device technologies by means of a Landauer-Büttiker-based extraction methodology (LBM) adapted for multiple 1D-channels. The extraction methodology yields values for an effective Schottky barrier height and a gate coupling coefficient, an indicator of the device working at the quantum capacitance limit. The novel LBM-based approach embracing the mechanisms in 1D electronics is compared to the conventional activation energy method not considering such effects. The latter approach underestimates the potential barrier height at metal-channel interfaces in comparison to the novel methodology. A test structure based on a displaced gate device is proposed based on numerical device simulation results towards an improved accuracy of the method. \vspace{1cm}\newline \footnotesize{\textit{This article may be downloaded for personal use only. Any other use requires prior permission of the author and AIP Publishing. This article appeared in Journal of Applied Physics and may be found at \href{https://aip.scitation.org/doi/10.1063/5.0091077}{DOI: 10.1063/5.0091077}}}
\end{abstract}

\maketitle

\preprint{ of the author and AIP Publishing.}
%
%
%
\section{Introduction}
\label{ch:intro}

One-dimensional (1D) field-effect transistor (FET) technologies embrace devices with quasi-ballistic materials and structures as the channel such as carbon nanotubes (CNTs) and nanowires (NWs). Single-tube and single-wire FETs have been useful to understand the corresponding device physics \cite{Kno21,WebMik17}. Transistors with an array of CNTs or NWs as channel have been demonstrated to be suitable candidates for practical low-power high-performance applications \cite{WuWu15,KilHel20,HilLau19,HarHer21}. The latter has been boosted by sophisticated techniques developed towards the integration of 1D-arrays-based devices in industry fabrication processes \cite{BisHil20,FriCer20,RayFri20,ShuPit14}.\fancyfoot[CO,CE]{And this is a fancy footer}

In contrast to single-1D-channel devices, multi-tube (MT) or multi-wire (MW) transistors present improved overall device characteristics, e.g., higher driving current capabilities and dynamic figures of merit \cite{ShuPit14,BraJin17,ChuWu19,KimLee13,MotSch18,DarMot19,HolPay10}. However, discussions on transport and injection phenomena in 1D FETs with an array of multiple channels are scarce in the literature. In order to improve 1D electronics technologies, a better understanding of the internal phenomena at metal-1D-channel interfaces within the same device, enabled by a reliable characterization, is required.

The interface characteristics at the metal contact regions and the 1D channel have a significant impact on the device performance. An important characteristic of this interface is the potential barrier height $\Phi_{\rm BH}$ dominated by a Schottky-like barrier height $\Phi_{\rm SB}$ (cf. Figs. \ref{fig:sym}(b), (c)). The latter is enabled by the different electronic properties between the tube/wire portions embedded within the metal contact and the uncoated 1D-channel \cite{Kno21,WebMik17,SveCam11,FedRyn16,ZhaCan12,FanKan16}. A weak Fermi level pinning at such interfaces prevents the evaluation of $\Phi_{\rm SB}$ with the Schottky-Mott approach \cite{SveCam11,FedRyn16,LeoTal06,FenHua16} and hence, extraction methods are required for its characterization.

The activation energy method (AEM), originally developed in the context of conventional semiconductor devices \cite{RodWil88}, has been used to obtain $\Phi_{\rm SB}$ values in multi-1D FETs in the literature \cite{GanLok17,KumNav19,XieZho21,DeMSac12,JeoBal17}. However, AEM does not cover the physics within the metal-1D-channel interface and it underestimates $\Phi_{\rm SB}$ in contrast to an extraction methodology embracing 1D transport as shown elsewhere for devices with single- \cite{PacCla17} and multi-1D channels \cite{PacRam20}. The latter methodology is identified as the 1D Landauer-Büttiker-equation based method (1D LBM) \cite{PacCla17}. 

In this work, 1D LBM is discussed in detail considering the transport in multi-1D-FETs in contrast to a previous work \cite{PacRam20}, where an interpretation of the underlying transport characteristics is missing. Contact interfaces of fabricated CNT- and NW-FET technologies are characterized with 1D-LBM here. Asymmetric gate devices are further investigated with a numerical device simulator in order to show the impact of the electrostatics on the accuracy of the extracted value.    

\section{Transport injection mechanisms and 1D-LBM in multi-1D FETs}



A phenomenological analysis of the transport in multi-1D-channel devices is given next by considering that \textit{(i)} screening effects due to tube/wire interactions are negligible and \textit{(ii)} there are not Schottky points within the channel. The first can be fulfilled in devices with a relaxed pitch \cite{MotSch18,DarMot19,HolPay10} whereas the latter is achieved in devices with tubes/wires properly aligned, a technology condition achievable for both CNTFET \cite{LiuHan20} and NWFET \cite{McIMor20} technologies.

\begin{figure}[!htb]
\centering
\includegraphics[height=0.183\textheight]{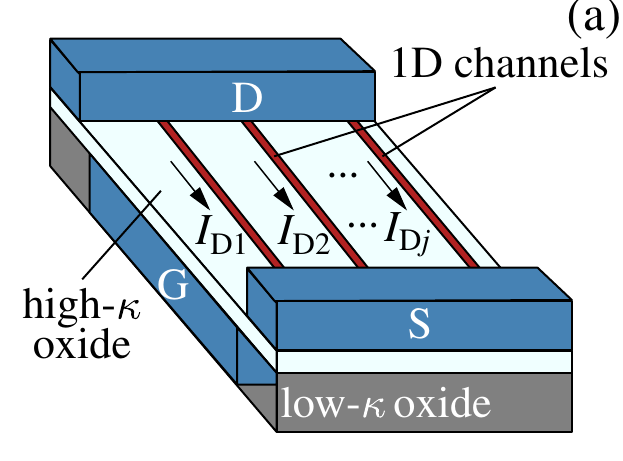}\\ \vspace{0.2cm}
\includegraphics[height=0.155\textheight]{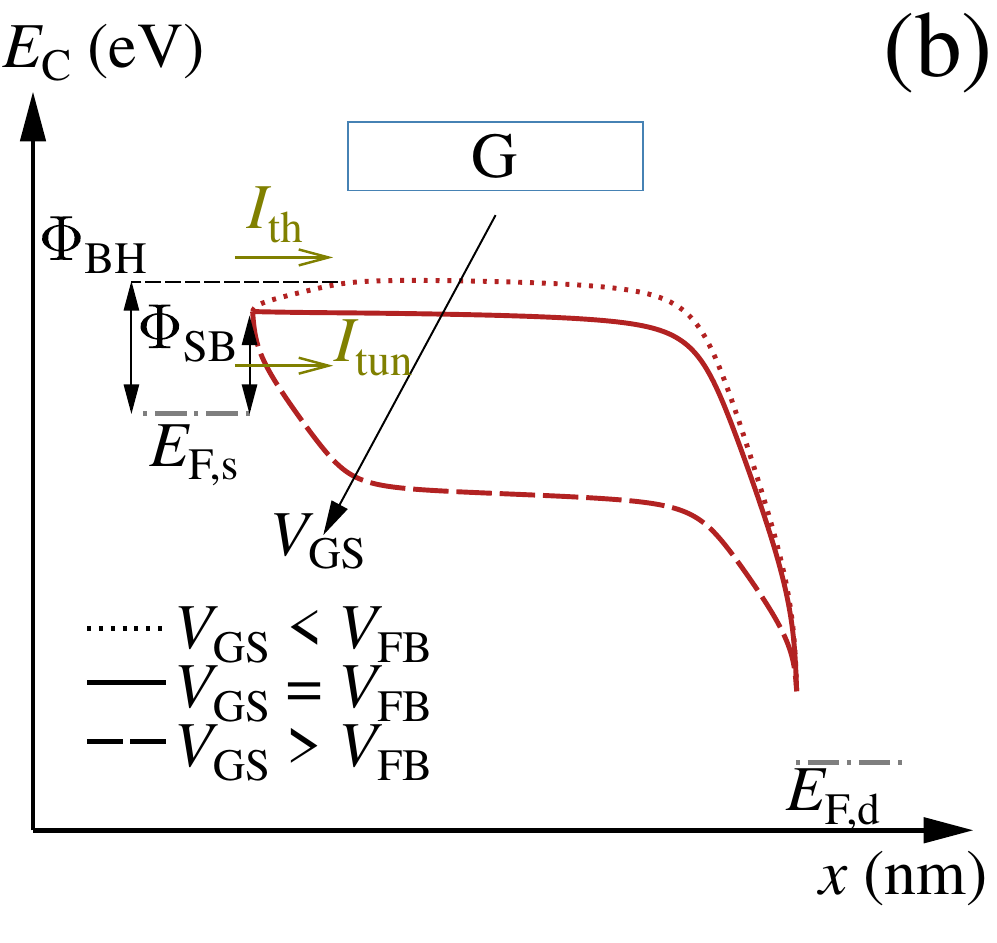}
\includegraphics[height=0.155\textheight]{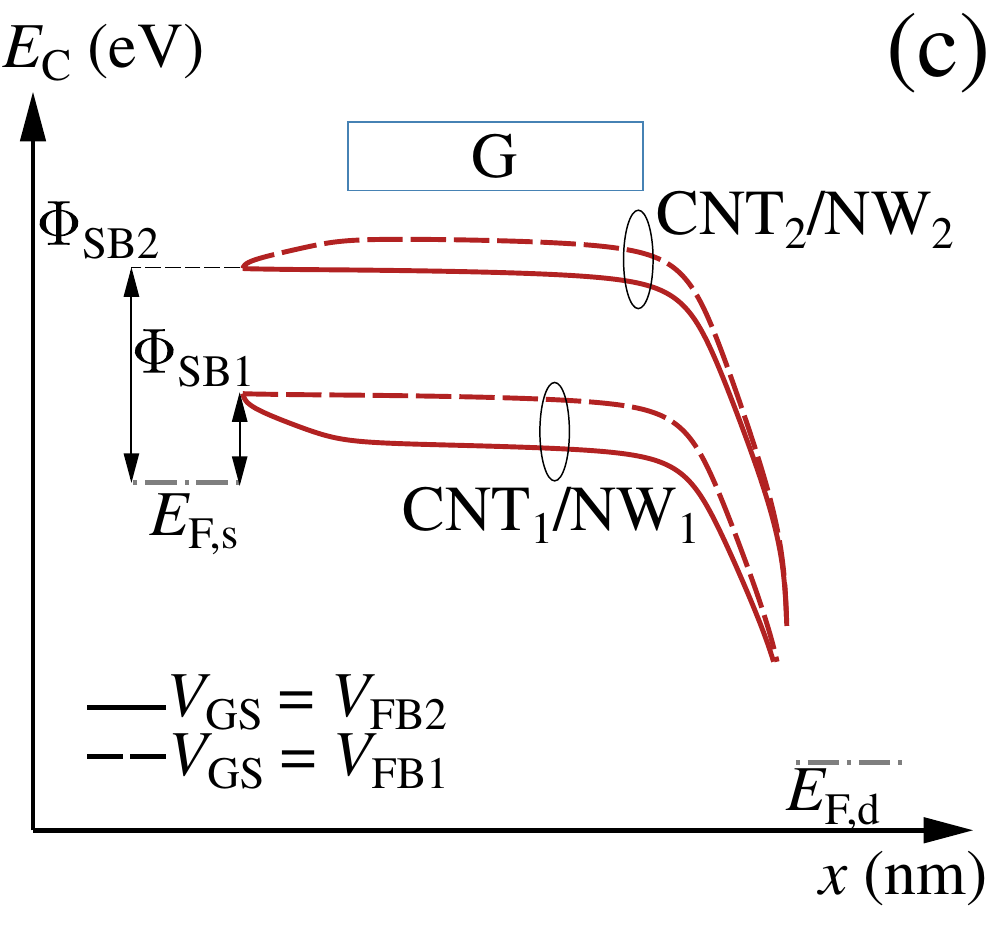}
\caption{(a) Schematic device structure (not at scale) of a buried gate multi-1D channel FET with perfectly aligned \textit{j}-tubes/wires. Sketch of the conduction band diagram at different $V_{\rm GS}$ for an \textit{n}-type multi-1D FET considering metal-tubes/wires interfaces with (b) identical and (c) different properties. Each pair of curves in (b) corresponds to one CNT/NW. A top-gate contact (G) is shown in (b) and (c) in order to ease the identification of the gated-channel region. $E_{\rm{F,s/d}}$ is the Fermi energy level at the source/drain contact. $V_{\rm FB}$ is a flat-band voltage and $I_{\rm th}$ and $I_{\rm tun}$ are the thermionic and tunneling current, respectively.}
\label{fig:sym}
\end{figure}

In aligned multi-CNT/NW-FETs, the carrier transport can be approximately described by a parallel network of quasi-ballistic channels as qualitatively depicted in Fig. \ref{fig:sym}(a) where the drain currents flowing through each channel $I_{\rm D,\it j}$ have been indicated. For devices with identical parallel channels, the overall transport injection mechanisms can be considered similar to the single-1D-channel Schottky FET, i.e., thermionic and tunneling transport are enabled by $\Phi_{\rm BH}$ before and after flat-band conditions have been reached as shown with the sketch of the conduction energy band in Fig. \ref{fig:sym}(b). In practice however, technological variations in the tubes/wires during the fabrication process, e.g., different tube/wire diameters, can lead to non-homogeneous metal-channel interfaces within the same device. The latter has been observed for both CNTFETs and NWFETs where tube/wire diameters impact on $\Phi_{\rm BH}$ \cite{KumNav19,CheApp05,SveSou09,WebHei14,HuWan20}. Hence, the individual flat-band conditions, and hence $\Phi_{\rm SB}$, differ among the 1D-channels in devices with non-identical contact interface properties. As observed in Fig. \ref{fig:sym}(c), a 1D-channel-related $\Phi_{\rm SB1}$ lower than other $\Phi_{\rm SB2}$ within the same device lead to pure thermionic injection for the whole device at the flat-band condition corresponding to the first channel (CNT$_1$/NW$_1$), however, tunneling injection is enabled for the first channel at the same bias at which only thermionic current ocurrs in the second channel (CNT$_2$/NW$_2$) due to its corresponding flat-band conditions. The characterization of $\Phi_{\rm SB}$ for each 1D-channel in multitube/multiwire is not trivial due to the different transport injection mechanisms discussed above. Nevertheless, for the overall device performance, the transport injection can be described by an effective $\Phi_{\rm BH}$ under certain considerations (cf. \textit{(i)} and \textit{(ii)} at the beginning of this section).

The total drain current $I_{\rm D}$ of multi-1D-FETs at the subthreshold regime can be approximately described as the sum of the individual quasi-ballistic current, given by the 1D-Landauer-Büttiker approach\cite{PacCla17,Cla11}, through each parallel path such as

\begin{equation}
\begin{aligned}
I_{\rm D} &= \sum_{j=1}^{\rm n_{\rm t/w}} I_{\rm D,\it j} \\ &\hspace{-0.2cm}\approx  \eta \sum_{j=1}^{\rm n_{\rm t/w}} \left\lbrace\exp\left[\frac{n_{\rm g,\it j}}{V_{\rm t}}\left(V_{\rm {GS}}-V_{\rm ot,\it j}\right)-\frac{\Phi_{\rm BH,eff}}{V_{\rm t}} + \frac{n_{\rm d,\it j}}{V_{\rm t}} V_{\rm DS}\right]\right\rbrace, 
\end{aligned}\label{eq:Id}
\end{equation}  

\noindent where $\rm n_{\rm t/w}$ is the total number of tubes/wires, $\eta=(4q^2/h)V_{\rm t}$ is a constant with $q$ as the electronic charge and $h$ as the Planck constant, $V_{\rm t}=k_{\rm B}T/q$ is the thermal voltage with $k_{\rm B}$ as the Boltzmann constant and $T$ the absolute temperature, $n_{\rm g}$ and $n_{\rm d}$ are gate and drain coupling coefficients \cite{Cla11}, respectively, $V_{\rm GS}$ and $V_{\rm DS}$ are the gate-to-source and drain-to-source voltage, and $\Phi_{\rm BH,eff}$ is an effective potential barrier height at which pure thermionic injection ocurrs in the multi-1D device. $V_{\rm ot}$ is the voltage indicating the onset of tunneling mechanisms which in single-tube/wire devices corresponds to the flatband voltage $V_{\rm FB}$ \cite{PacCla17}. Eq. (\ref{eq:Id}) has been obtained by considering that the thermionic transmission probability $\mathcal{T}$ is equal to one as expected for sub-\SI{}{\micro\meter} CNTFETs and NWFETs, specially at low electric fields \cite{BraJin17,ChuWu19,KarSch16,TanSud18,XuQiu19}. An energy difference of $E_{\rm cc}-E \geq 3k_{\rm B}T$ has been considered as well, with $E_{\rm cc}$ as the current control energy \cite{Cla11,Pac19} at which transport is enabled. For electron/hole transport, $E_{\rm cc}$ is generally associated to the minimum/maximum of the conduction/valence band within the gated-channel region. After rearranging Eq. (\ref{eq:Id}), an expression for $\Phi_{\rm BH,eff}$ yields \cite{PacCla17,Pac19}

\begin{equation}
\Phi_{\rm BH,eff} \approx \sum_{j=1}^{\rm n_{\rm t/w}} \left[-\frac{k_{\rm B}}{q}\alpha_{j} + n_{\rm g,\it j}\left(V_{\rm {GS}}-V_{\rm ot,\it j}\right) + n_{\rm d,\it j} V_{\rm DS}\right],
\label{eq:phi_eff}
\end{equation}

\noindent where the term $\alpha$ corresponds to the slope of the Arrhenius plots of $\ln\left(I_{\rm D}/T\right)$ versus $1/T$. Eq. (\ref{eq:phi_eff}) reveals a linear relation between $\Phi_{\rm BH,eff}$ and a term embracing a temperature- and drain-induced-electrostatics-dependent potential step $\Phi_{\rm SB,eff}$, identified as an effective Schottky-like barrier height, such as 

\begin{equation}
\Phi_{\rm BH,eff} \approx \sum_{j=1}^{\rm n_{\rm t/w}} \left[ \Phi_{\rm SB,eff} + n_{\rm g,\it j}\left(V_{\rm {GS}}-V_{\rm ot,\it j}\right)\right],
\label{eq:phi_eff2}
\end{equation}
 
\noindent from which $\Phi_{\rm SB,eff}$ can be extracted at $\vert V_{\rm GS} = \min (V_{\rm ot,\it j}) \vert$. A $V_{\rm GS}$-dependent plot of Eq. (\ref{eq:phi_eff}), valid only for pure thermionic transport, enables the identification of $\vert \min (V_{\rm ot,\it j})\vert$ at a $V_{\rm GS}$ where the linear behavior of $\Phi_{\rm BH,eff}$ vanishes. In contrast to AEM, 1D LBM enables the extraction of the terms $n_{\rm d}$ and $n_{\rm g}$, from $\alpha$ and $\partial \left(\ln I_{\rm D}\right) / \partial V_{\rm GS}$, respectively. The extraction of $n_{\rm d}$ is enabled only if different $V_{\rm DS}$ are evaluated. Notice that the individual characteristics of each 1D-channel required to compute the above equations, e.g., $\alpha$ for each tube/wire, are challenging to obtain in practice, however, experimental data embraces the total contribution of all channels, i.e., the entire sum terms in Eqs. (\ref{eq:Id})-(\ref{eq:phi_eff2}), and hence, the extraction method (1D LBM) can be applied. The methodology is illustrated next with data of fabricated CNTFETs and NWFETs.

For long devices, the extraction method is justified by considering a weak temperature dependence of the scattering mechanisms at low-fields, i.e., $\mathcal{T}$ is considered a constant value. The latter enables to eliminate $\mathcal{T}$ when obtaining $\alpha$ and, subsequently, Eqs. (\ref{eq:phi_eff}) and (\ref{eq:phi_eff2}). 

\section{Results across CNTFET and NWFET technologies}

\subsection{Extraction from experimental data}

The 1D LBM has been applied to available experimental data in the literature of multi-channel CNTFETs \cite{GanLok17,XieZho21,BroRot18,SriHil19,GotSch21} and NWFETs \cite{ChuWu19,JeoBal17,JeoPre15} from different technologies, i.e., different channel lengths and gate architectures. Additionally, a \SI{1.5}{\micro\meter}-long NWFET, labeled as NWFET$_{\rm EPFL}$, not presented before has been also characterized. The fabrication process for NWFET$_{\rm EPFL}$ has been described elsewhere\cite{FriCer18}. Device geometry parameters of the devices under study are summarized in Table \ref{tab:geom} for reference purposes.

\begin{table} [!htb] 
\begin{center}
\caption{Device characteristics (channel/gate length $L_{\rm ch/g}$, CNT/NW diameter $d$, CNT/NW density $D$, equivalent oxide thickness $EOT$) and gate architectures (global-back-gate GBG, buried-back-gate BG, top-gate TG, gate-all-around GAA) of fabricated multi-channel 1D devices. Missing data are indicated with a --.}
\begin{tabular}{c|c|c|c|c|c|c|c}

device & ref. & \makecell{gate\\arch.} & \makecell{$L_{\rm ch}$\\(\SI{}{\micro\meter})} & \makecell{$L_{\rm g}$\\(\SI{}{\micro\meter})} & \makecell{$d$\\(\SI{}{\nano\meter})} & \makecell{$D$\\(\SI{}{\micro\meter^{-1}})} & \makecell{$EOT$\\(\SI{}{\nano\meter})} \\ \hline

\multirow{5}{*}{CNTFET} & \cite{SriHil19} & BG & \SI{0.5}{} & \SI{0.74}{} & -- & -- & \SI{2.5}{} \\
& \cite{GanLok17} & GBG & \SI{1}{} & \SI{1}{} & \SI{1.4}{} & -- & \SI{290}{}\\
& \cite{BroRot18} & TG & $\SI{24}{}$ & -- & \SI{0.76}{} & $\geq\SI{40}{}$ & $\sim\SI{280}{}$\\
& \cite{BroRot18} & TG & $\SI{24}{}$ & -- & \SIrange{0.76}{1.31}{} & $\geq\SI{40}{}$ & $\sim\SI{280}{}$\\
& \cite{XieZho21} & TG & -- & \SI{2}{} & \SI{1.3}{} & -- & -- \\ 
& \cite{GotSch21} & TG & $\SI{40}{}$ & -- & \SI{0.76}{} & \SI{15}{} &  $\sim\SI{280}{}$ \\ \hline

\multirow{3}{*}{NWFET} & \cite{ChuWu19}  & TG & \SI{0.05}{} & -- & \SI{40}{}  & \SI{5.6}{} & \SI{5}{} \\

& \cite{JeoBal17} & GAA & \SI{1}{} & \SI{1}{} & \SI{20}{} & \SI{12}{} & \SI{46}{}\\

& \cite{JeoPre15}  & GAA & \SI{6}{} & \SI{1}{} & \SI{20}{} & \SI{12}{} & \SI{46}{}\\

& $_{\rm EPFL}$ & GBG & \SI{1.5}{} & \SI{1.5}{} & \SI{20}{} & -- & $\sim\SI{25}{}$\\

\end{tabular} \label{tab:geom}
\end{center}
\end{table}

\begin{figure}[!htb]
\centering
\includegraphics[height=0.1685\textheight]{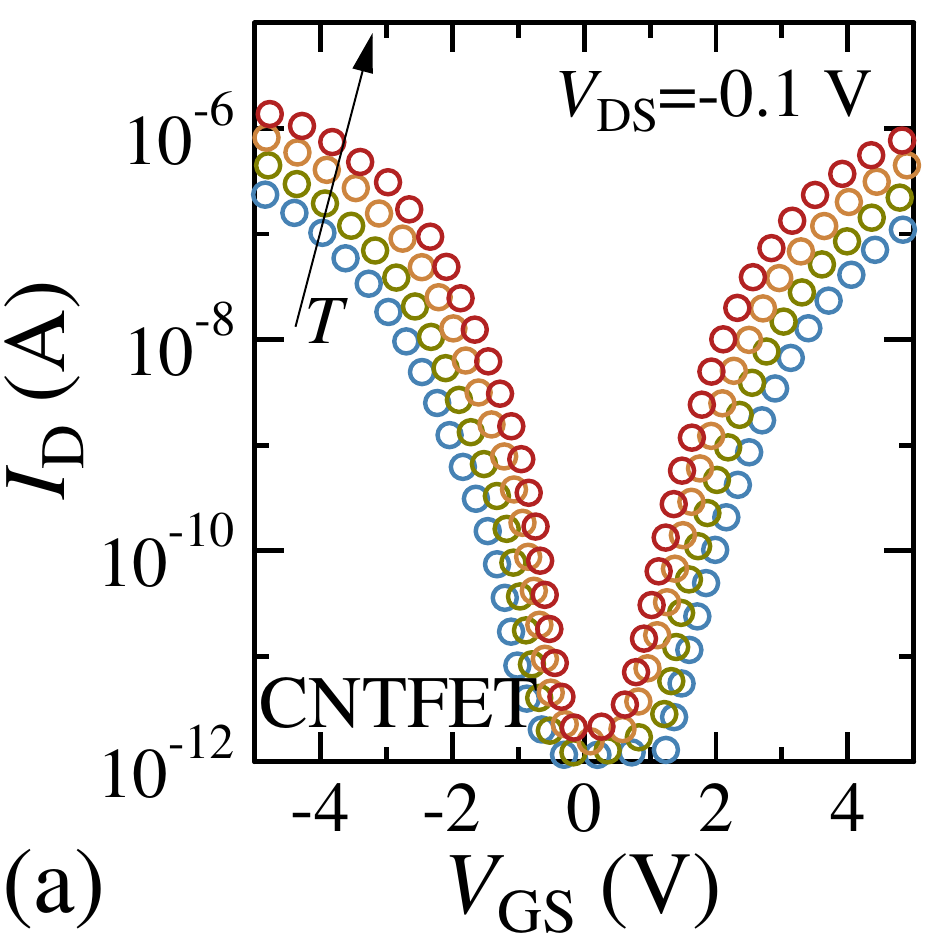}
\includegraphics[height=0.1685\textheight]{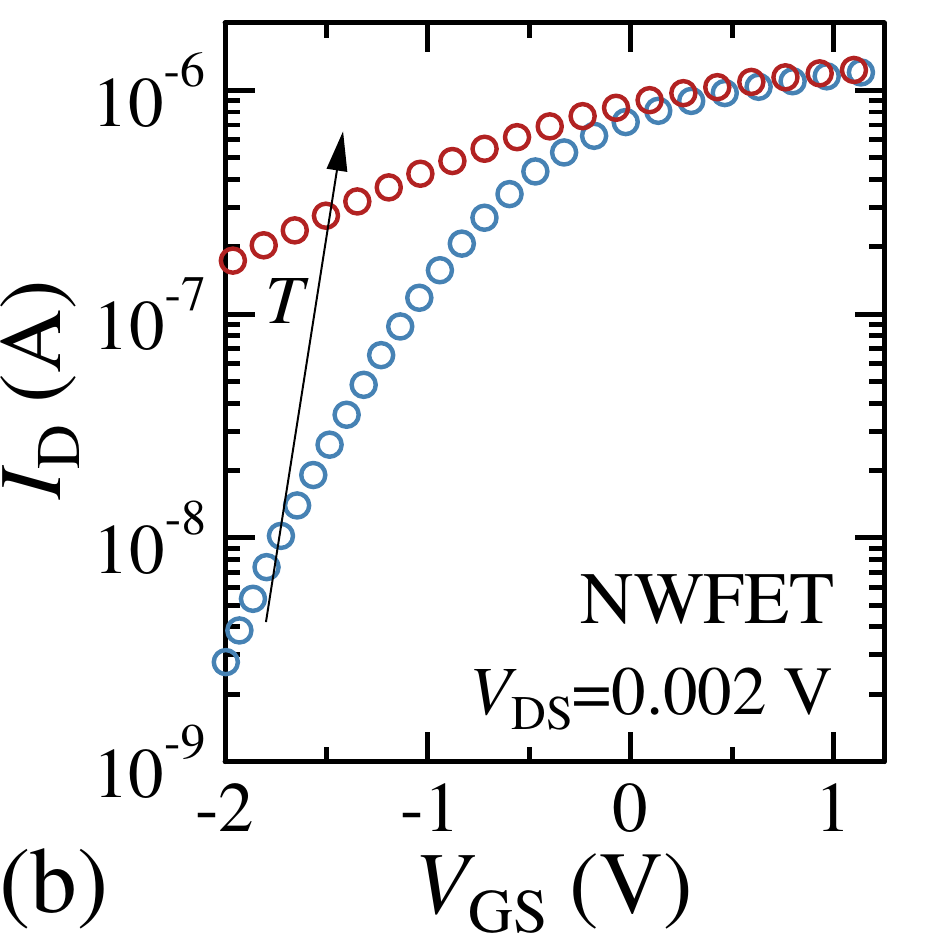} \\
\includegraphics[height=0.1685\textheight]{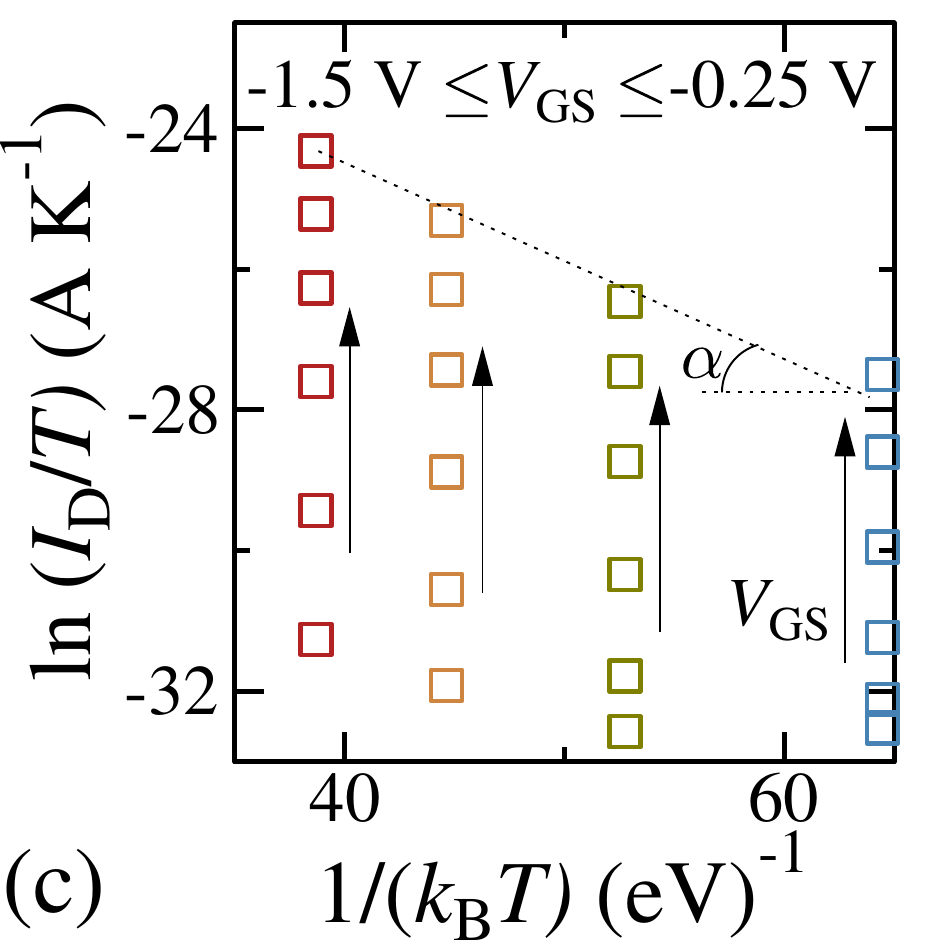}
\includegraphics[height=0.1685\textheight]{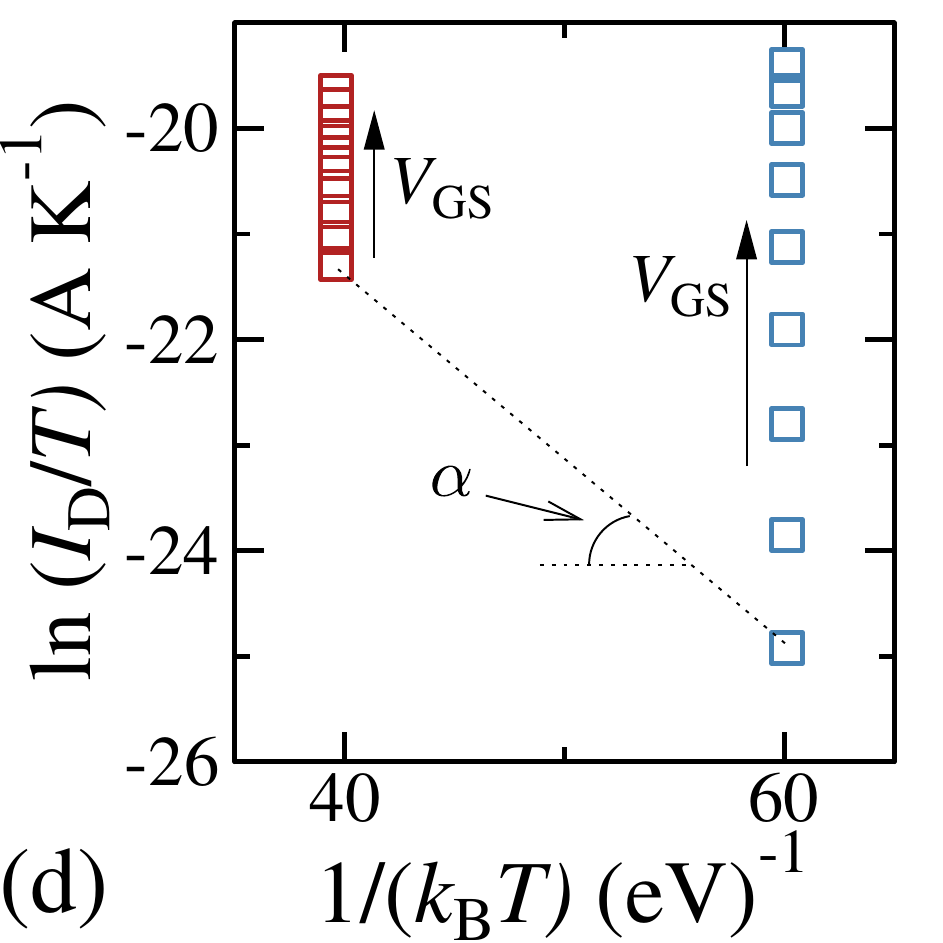} \\
\includegraphics[height=0.1685\textheight]{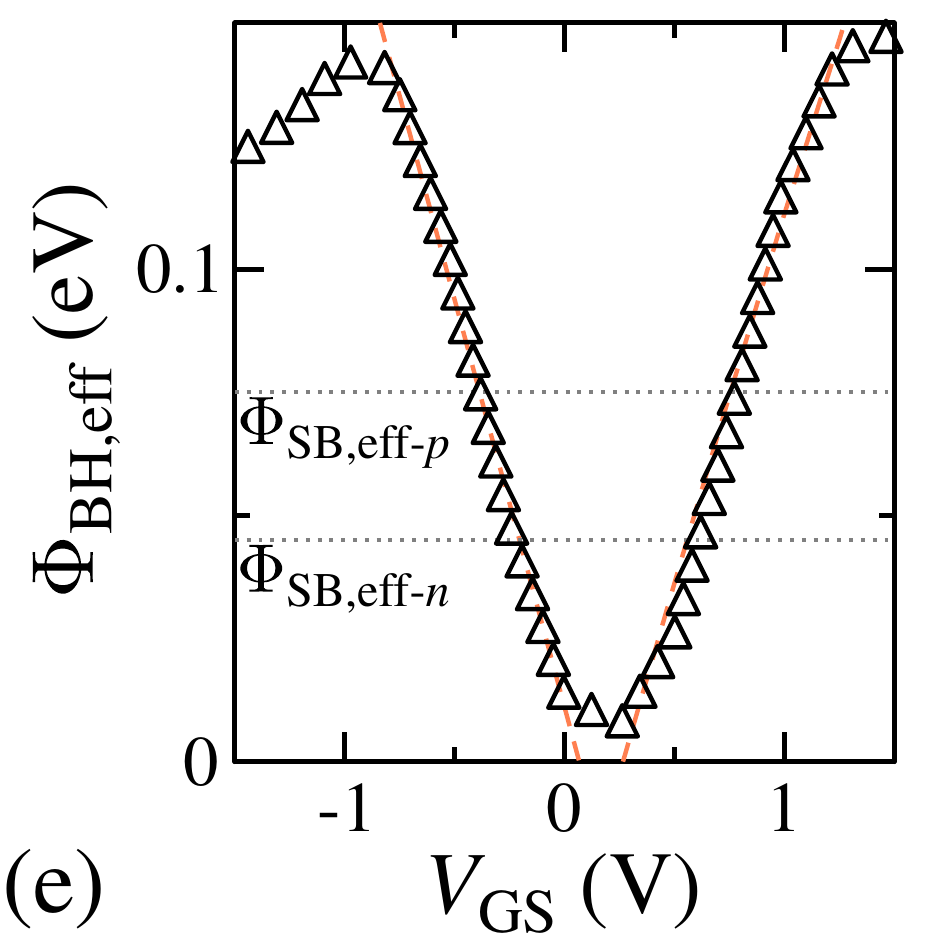}
\includegraphics[height=0.1685\textheight]{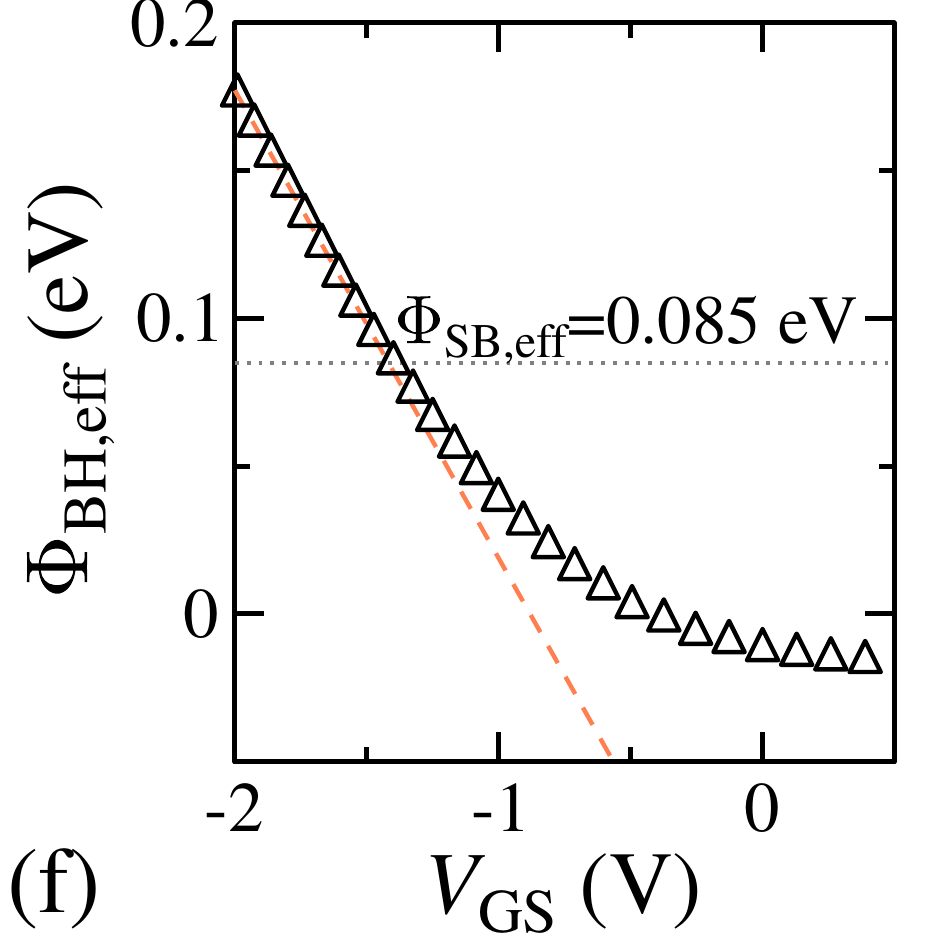} \\
\caption{Extraction of effective Schottky barrier height of (a), (c), (e) a multitube CNTFET \cite{BroRot18} and (b), (d), (f) a multiwire NWFET. Transfer characteristics at different temperatures for the (a) CNTFET ($T$=\SIlist{180;220;260;300}{\kelvin}) and (b) NWFET ($T$=\SIlist{193;293}{\kelvin}). (c)-(d) Arrhenius plot at the subthreshold regime; dotted lines are added as a guide for the eyes in order to show the extraction of $\alpha$ at each $V_{\rm GS}$. (e)-(f) $\Phi_{\rm BH,eff}$ over $V_{\rm GS}$ from which the effective Schottky barrier height (for electrons or holes) is extracted; dashed lines are a linear extrapolation and dotted lines are added as a guide for the eyes in order to indicate the extracted values.}
\label{fig:extr}
\end{figure}

Fig. \ref{fig:extr} shows the extraction of $\Phi_{\rm SB,eff}$ of the \SI{24}{\micro\meter}-long multitube CNTFET\cite{BroRot18} with tube diameters of $\sim$\SI{0.76}{\nano\meter} as well as of the \SI{1.5}{\micro\meter}-long NWFET$_{\rm EPFL}$. The transfer characteristics of both devices at different temperature and at a specific $V_{\rm DS}$ are shown in Figs. \ref{fig:extr}(a) and (b) where the asymmetric ambipolarity of the CNTFET can be observed. The term $\alpha$ has been extracted from the Arrhenius plots of each device shown in Figs. \ref{fig:extr}(c) and (d) at $V_{\rm GS}$ within the subthreshold regime where Eq. (\ref{eq:Id}) is valid. The more temperatures available, the more accurate the extraction of $\alpha$ is. By obtaining $\alpha$ at each $V_{\rm GS}$, the plots of $\Phi_{\rm BH,eff}$ shown in Figs. \ref{fig:extr}(e) and (f) have been obtained from which the onset of tunneling processes at a $V_{\rm ot}$ is identified. In this work, a relative error of $\sim$\SI{0.5}{}\% between the linear extrapolation and $\Phi_{\rm BH,eff}$ is considered in order to identify $V_{\rm ot}$. The effective Schottky barrier heights are identified at $V_{\rm ot}$. For the ambipolar CNTFET, both Schottky barriers for electrons and holes can be identified by considering the bias range in which thermionic emission from each type of carrier is expected to be dominant.


\begin{figure}[!htb]
\centering
\includegraphics[height=0.1685\textheight]{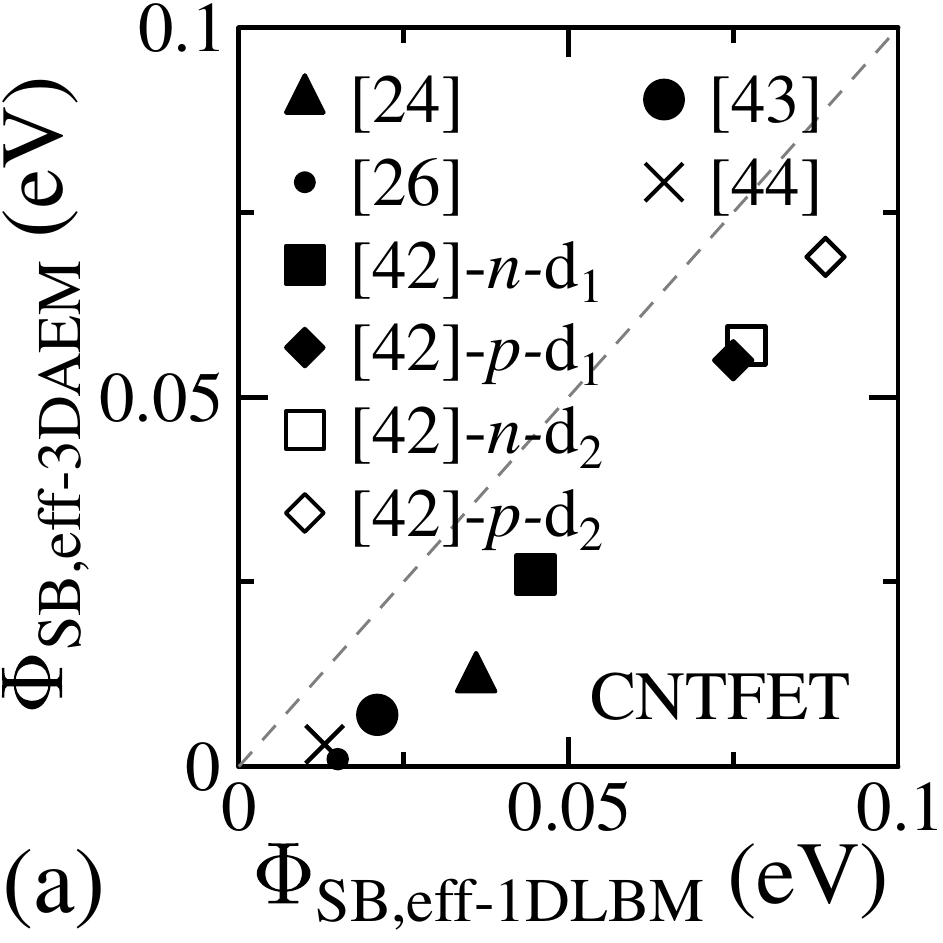}
\includegraphics[height=0.1685\textheight]{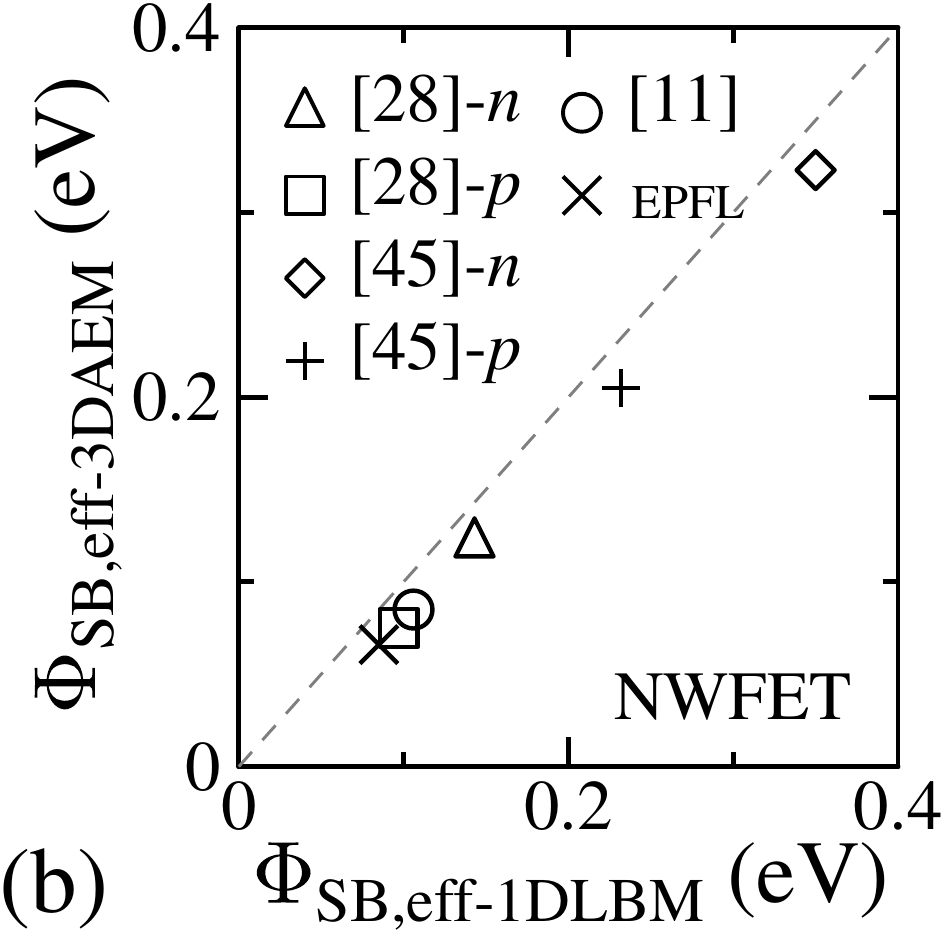}
\caption{Extracted effective Schottky barrier height with 1D LBM and 3D AEM of fabricated (a) CNTFETs and (b) NWFETs. Dashed lines are added as a guide for the eye and they separate bottom and upper regions where the extracted 1D LBM values are higher or lower, respectively, than the 3D AEM values.}
\label{fig:phi}
\end{figure}

The extracted 1D LBM $\Phi_{\rm SB,eff}$s of the CNTFETs \cite{GanLok17,XieZho21,BroRot18,SriHil19,GotSch21} and NWFETs \cite{ChuWu19,JeoBal17,JeoPre15} under study, including NWFET$_{\rm EPFL}$, have been compared in Fig. \ref{fig:phi} with the values obtained with the conventional AEM considering a three dimensional system. Notice that for the ambipolar devices \cite{JeoBal17,JeoPre15} including the CNTFETs with identical device geometry but different CNT diameters distribution reported in \cite{BroRot18}, $\Phi_{\rm SB,eff}$ values have been extracted for both \textit{n}-type and \textit{p}-type transport. 

In all devices under study, $\Phi_{\rm SB,eff-1DLBM}$ is higher than $\Phi_{\rm SB,eff-3DAEM}$, similar to the case of single-1D-channel devices \cite{PacCla17}. The underestimated 3D AEM values with respect to the 1D LBM approach can be explained by \textit{(i)} a dimensionality issue leading to a $T^2$ factor yielded from a 3D-system consideration in the Arrhenius function rather than the $T$ factor related to 1D interfaces (e.g., see definition of $\alpha$ in Eq. (\ref{eq:phi_eff})) and \textit{(ii)} neglected gate and drain coupling coefficients in the underlying equation used in AEM \cite{Pac19}. Further details on the difference between the methods are provided in the Appendix. The higher accuracy of 1D LBM with respect to 3D AEM has been previously demonstrated for single-channel devices \cite{PacCla17}. 

For the CNTFETs with identical device architectures but different diameter distributions \cite{BroRot18}, $\Phi_{\rm SB,eff-1DLBM}$ extracted values are higher (regardless the type of transport) for the device with the largest tube diameter distribution, similarly to the ones obtained by an AEM-like  method for these same transistors \cite{BroRot18} but in contrast to previous findings in single-tube transistors \cite{CheApp05,SveSou09}. This contradiction with the one-channel case has been associated to thermionic injection hindered by tunneling mechanisms in non-homogeneous interfaces in multi-1D channel devices \cite{PacRam20}. Furthermore, non-homogeneous channel bands due to tube crossings might also impact the extraction as discussed below (cf. section III.C).

\begin{figure}[!htb]
\centering
\includegraphics[height=0.1685\textheight]{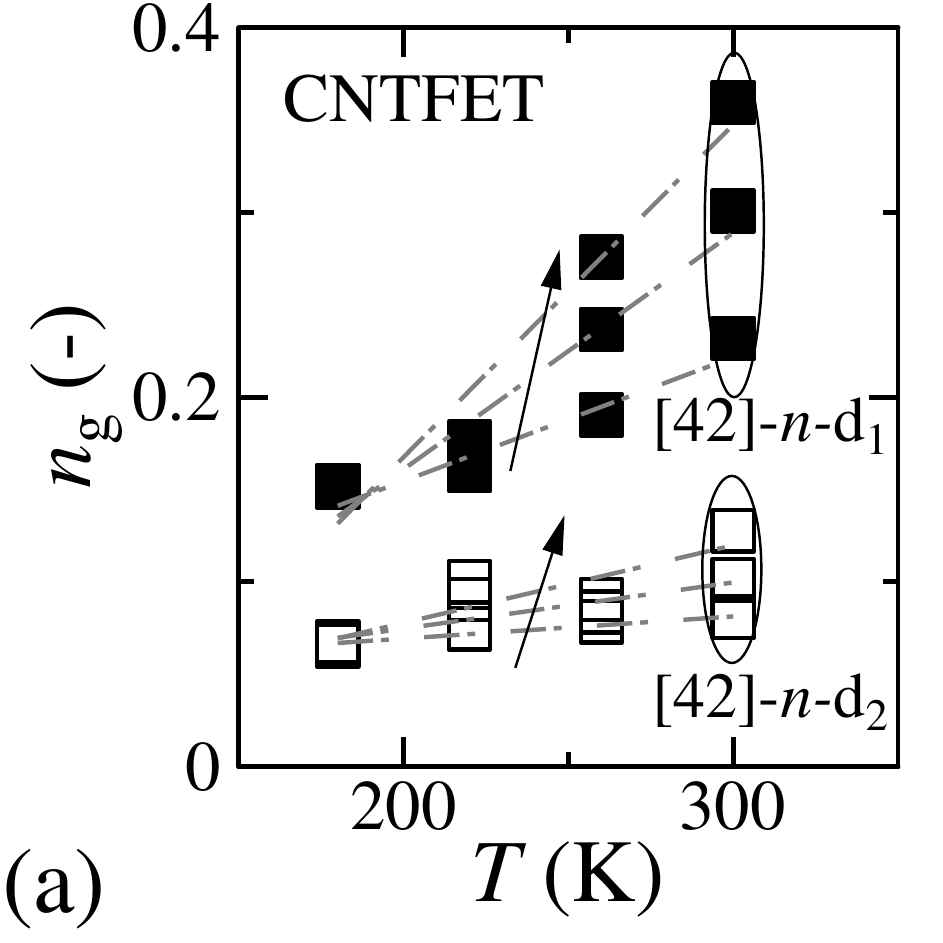}
\includegraphics[height=0.1685\textheight]{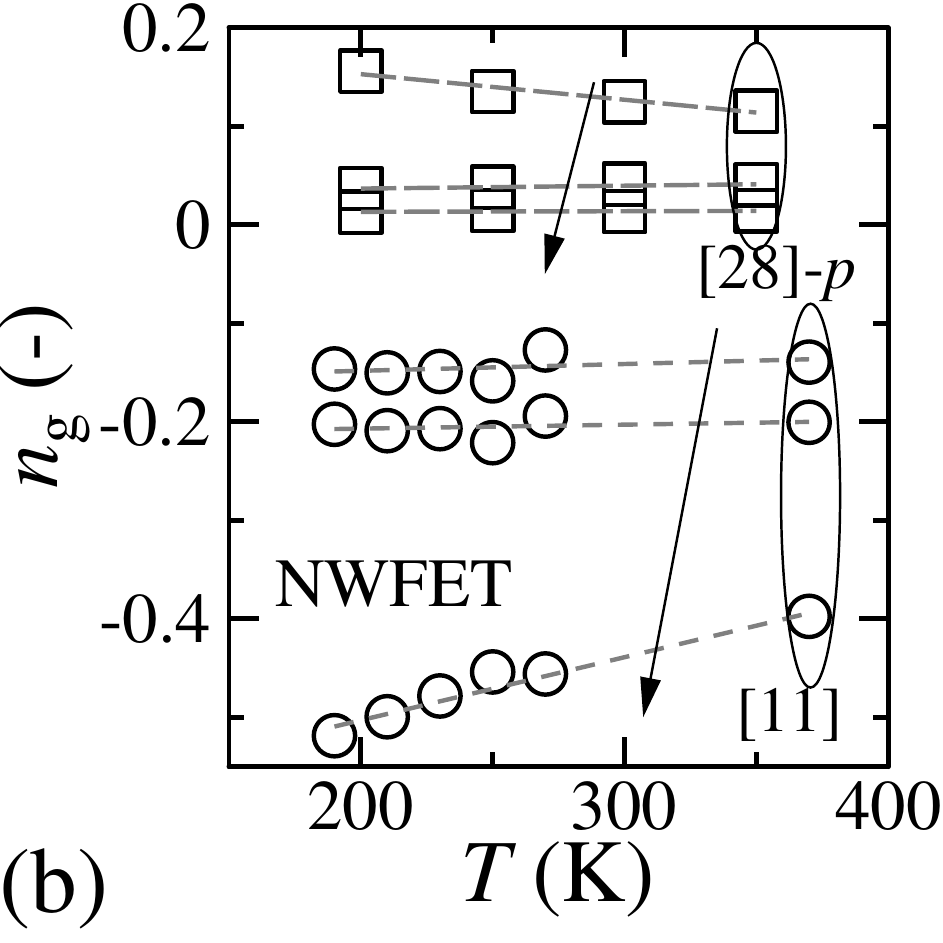}
\caption{Extracted gate couple coefficient over temperature of (a) CNTFETs \cite{BroRot18} and (b) NWFETs \cite{ChuWu19,JeoBal17}. Dashed lines are added as a guide for the eye and they indicate trends. Solid lines with arrow indicate an increasing $V_{\rm GS}$ for the different data sets.}
\label{fig:ng}
\end{figure}

In contrast to AEM, 1D LBM enables the extraction of a device high-performance indicator such as the gate coupling coefficient: $\vert n_g \vert \rightarrow 1$ indicates an operation regime known as the quantum capacitance limit \cite{MotCla15,RazJan13}. Fig. \ref{fig:ng} shows the extracted $\vert n_{\rm g} \vert$ of some of the devices under study evaluated at different $T$ and $V_{\rm GS}$ (within the bias region of validity of the method). For the \SI{24}{\micro\meter}-long CNTFETs \cite{BroRot18}, the smaller the diameter the better the control of the gate over the channel is, and hence, it leads to a steeper subthreshold slope as suggested elsewhere \cite{PacFuc18} and observed in the experimental transfer characteristics of these devices\cite{BroRot18}. The highest $\vert n_{\rm g} \vert$ extracted for the NWFET studied here has been achieved for the shortest device \cite{ChuWu19}, due to the thin \textit{EOT} (cf. Table I), i.e., the corresponding high gate capacitance dominates over the wire capacitance. Interestingly, an increasing temperature improves $n_{\rm g}$ for the CNTFETs while the contrary is observed for the NWFETs. Since the latter can ocurr due to a trade-off between thermal-dependent phenomena, e.g., contact resistance, scattering mechanisms, etc., a further analysis can be suggested for CNT/NW devices with similar gate and channel architectures, however, this is out of the scope of the present study. 

The onset of tunneling mechanisms in the $\Phi_{\rm BH}(V_{\rm GS})$-plot where $\Phi_{\rm SB}=\Phi_{\rm BH}\vert_{V_{\rm GS}=V_{\rm FB}}$ can be challenging to identify in devices where \textit{(i)} $n_{\rm g}$ has a weak $V_{\rm GS}$-dependence \cite{SanBla18} or \textit{(ii)} additional apparent linear regions of such plot are enabled by transparent contacts \cite{PacCla17}. In order to overcome these challenges, a special test structure to enhance the $\Phi_{\rm SB,eff}$-extraction by 1D-LBM is suggested next for multi-1D-channel devices.


\subsection{Test structure proposal}
%
%
%

\begin{figure}[!htb]
\centering
\includegraphics[height=0.10\textheight]{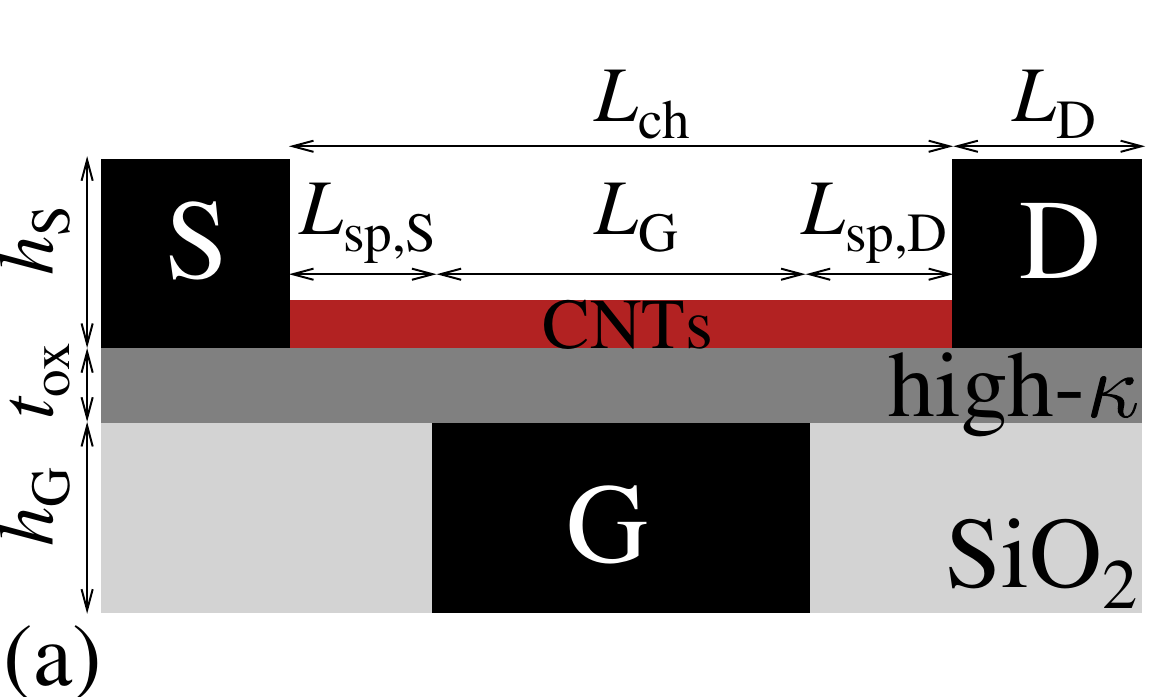}
\includegraphics[height=0.10\textheight]{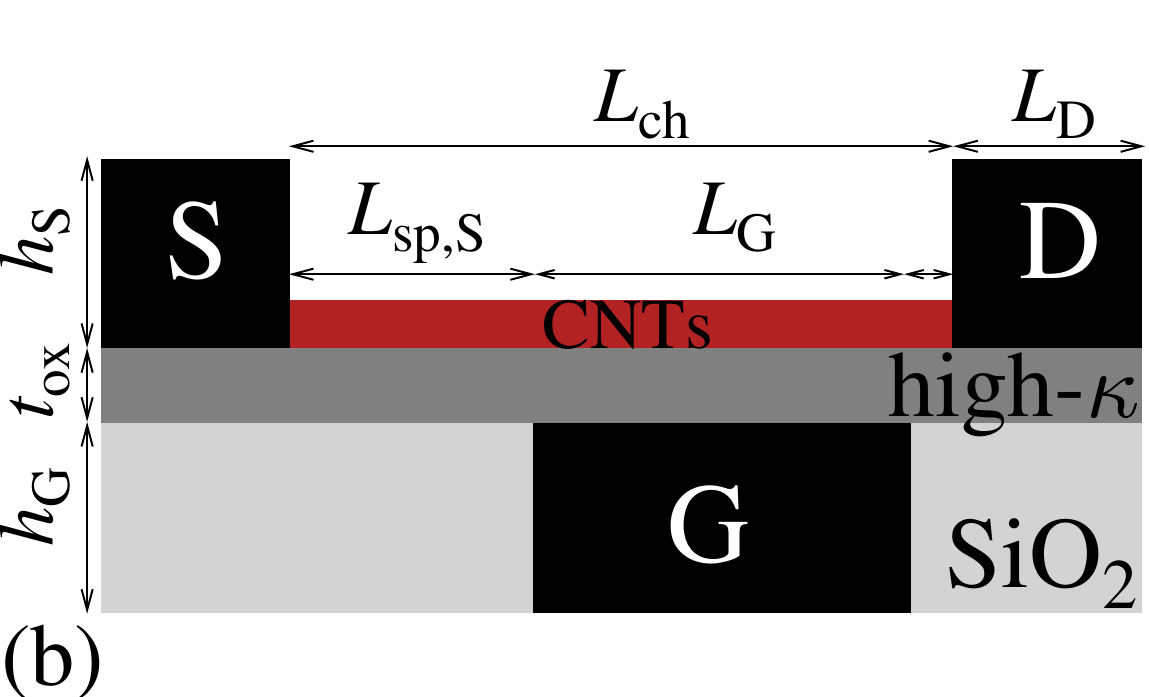}
\caption{Schematic cross sections (not drawn to scale) of simulated CNTFETs with (a) a symmetric buried gate (MTBG1) and (b) an asymmetric buried gate (MTBG2). Gate oxide has a permittivity of \SI{16}{}. For both devices $h_{\rm G}=\SI{200}{\nano\meter}$, $h_{\rm S/D}=\SI{100}{\nano\meter}$, $t_{\rm ox}=\SI{15}{\nano\meter}$, $L_{\rm S/D}=\SI{50}{\nano\meter}$, $L_{\rm G}=\SI{200}{\nano\meter}$ and $L_{\rm ch}=\SI{500}{\nano\meter}$ where as $L_{\rm sp,S}$ and $L_{\rm sp,D}$ are of \SI{150}{\nano\meter} both for (a) and of \SI{250}{\nano\meter} and \SI{50}{\nano\meter} for (b), respectively.}
\label{fig:sim}
\end{figure}

An experimentally-verified in-house numerical CNTFET simulator using a self-consistent solution of a transport equation and the Poisson equation, presented elsewhere \cite{MotSch18,ClaMot14,Mot19}, has been used here in order to propose a test structure for improving the $\Phi_{\rm SB,eff}$ extraction. Two \textit{n}-type multi-tube (MT) BG CNTFETs with identical device architectures and channel characteristics but different spacer lengths $L_{\rm sp,x}$ have been studied. The latter architecture diminishes tunneling mechanisms \cite{LiuHan20,SriHil19} and hence eases the extraction. Schematic cross sections of the simulated devices are shown in Fig. \ref{fig:sim}. The metal-CNT interfaces in each device are a combination of the practical case of three parallel non-homogenous tubes within the device channel \cite{PacRam20} enabling a different Schottky barrier height each (cf. Fig. \ref{fig:sym}(c)) \cite{CheApp05,SveSou09}: \SIlist{0.05;0.1;0.2}{\electronvolt}. The device width is \SI{60}{\nano\meter} yielding a tube density of \SI{50}{} CNT \SI{}{\micro\meter^{-1}} associated to negligible screening effects between tubes \cite{MotSch18}. 

%
%
%
%
%

\begin{figure}[!htb]
\centering
\includegraphics[height=0.1685\textheight]{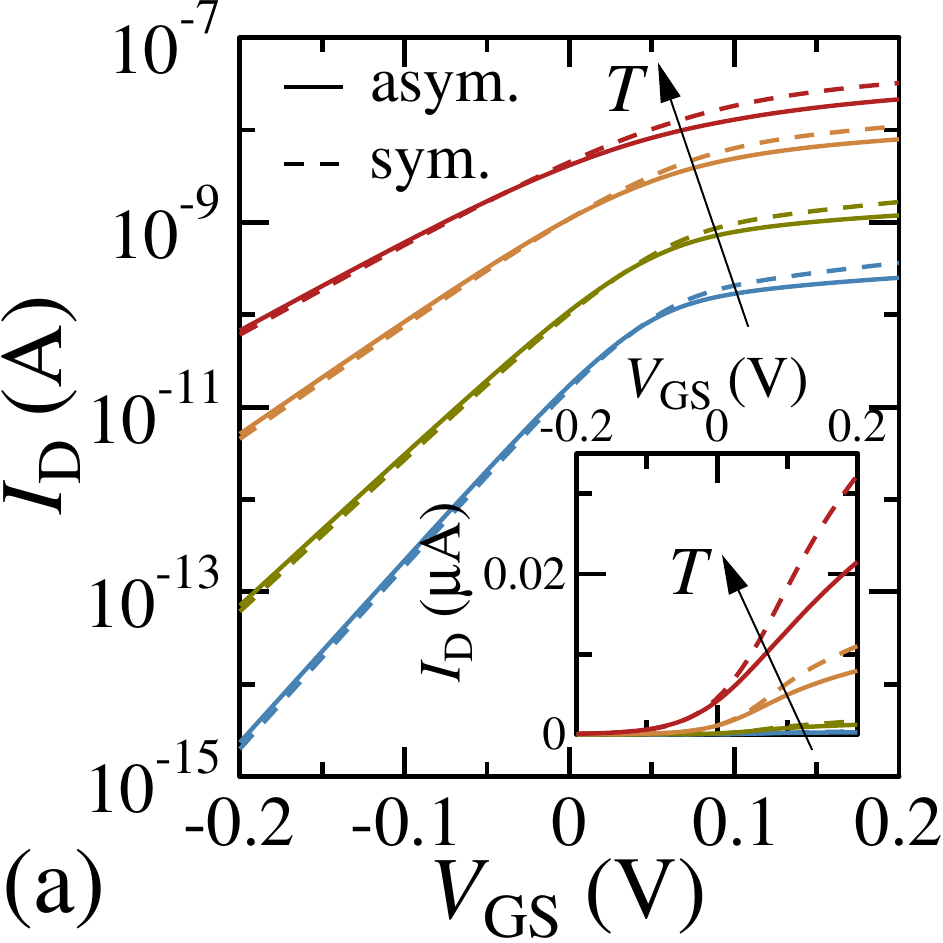}
\includegraphics[height=0.1685\textheight]{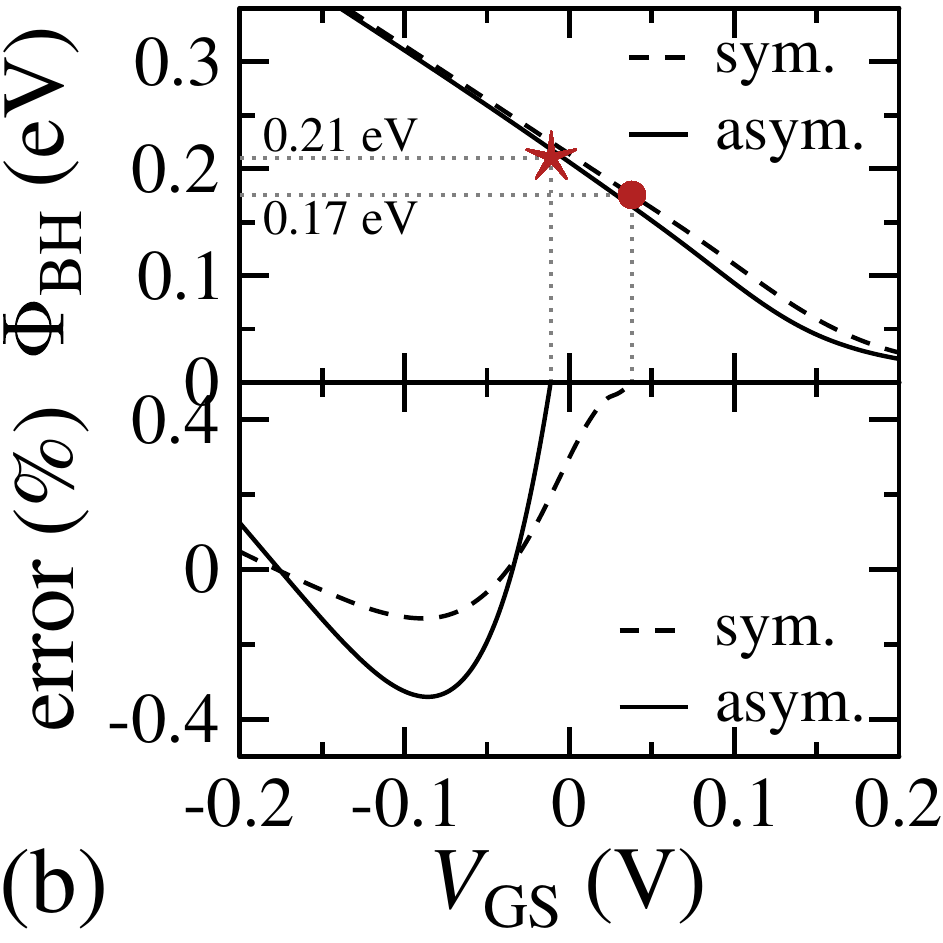}
\caption{Data of simulated symmetric and asymmetric BG MT CNTFETs. (a) Transfer characteristics at $V_{\rm DS}$=\SI{0.2}{\volt} and different temperature. (b) Schottky barrier height extraction from the barrier height potential plot over $V_{\rm GS}$ (top) and the relative error of such plot related to a linear extrapolation of the pure thermionic response (bottom). Red star (red dot) indicates the point at which $\Phi_{\rm SB,eff}$ has been extracted for the (a) symmetric and (b) asymmetric device.}
\label{fig:coos_results}
\end{figure}

Simulations have been performed at \SIlist{250;300;400;500}{\kelvin} and $V_{\rm DS}$ equal to \SI{0.2}{\volt} over a $V_{\rm GS}$ range between \SIrange{-0.2}{0.2}{\volt}. Tunneling and scattering mechanisms have been both considered whereas only electron transport has been enabled for simplification purposes. The simulated transfer characteristics at different temperatures of both MT CNTFETs are shown in Fig. \ref{fig:coos_results}(a). The 1D-LBM applied to the simulation data yields $\Phi_{\rm SB,eff}$s of \SI{0.17}{\electronvolt} and \SI{0.21}{\electronvolt} for the symmetric and asymmetric gate device, respectively, as shown in Fig. \ref{fig:coos_results}(b) where a relative error of $\approx$\SI{0.5}{}\% between a linear trend and the $\Phi_{\rm BH} (V_{\rm GS})$-plot has been used to identify an extracted flat-band voltage $V_{\rm FB,ext}(=V_{\rm ot,ext})$ and, consequently, to extract $\Phi_{\rm SB,ext}$ for each case. $V_{\rm FB,ext}$ are equal to \SI{0.037}{\volt} and -\SI{0.007}{\volt} for the symmetric and asymmetric devices, respectively.

\begin{figure}[!htb]
\centering
\includegraphics[height=0.1685\textheight]{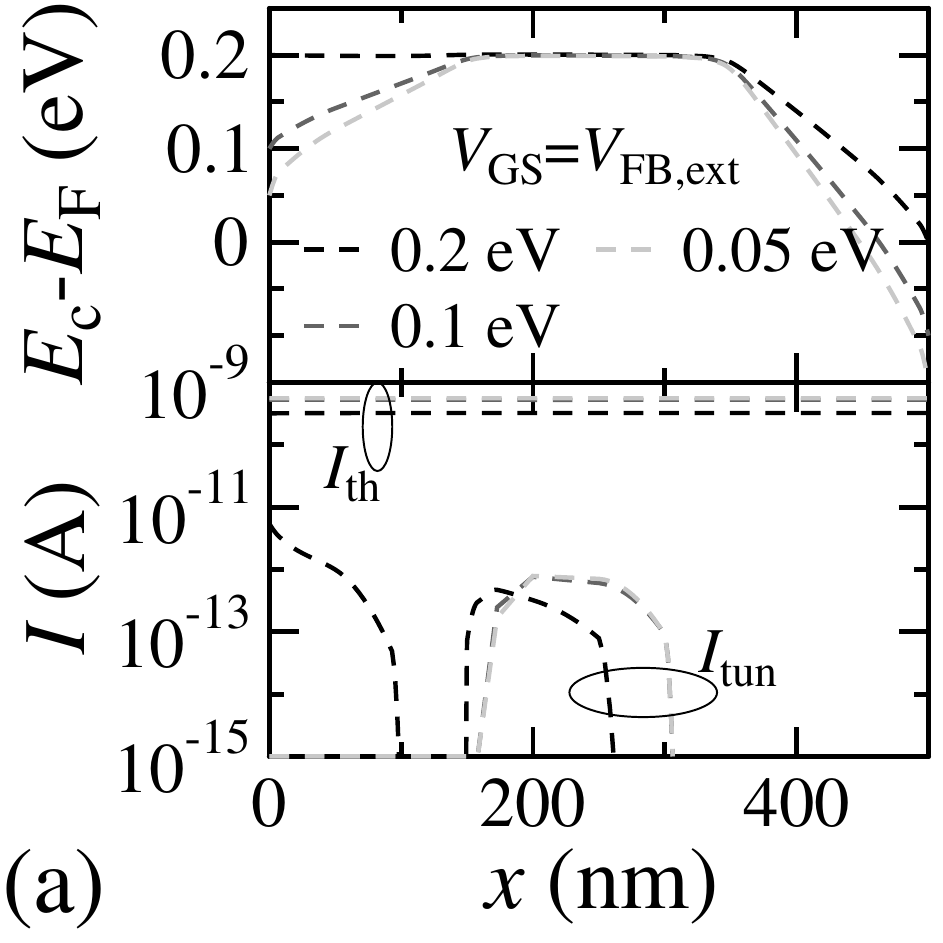}
\includegraphics[height=0.1685\textheight]{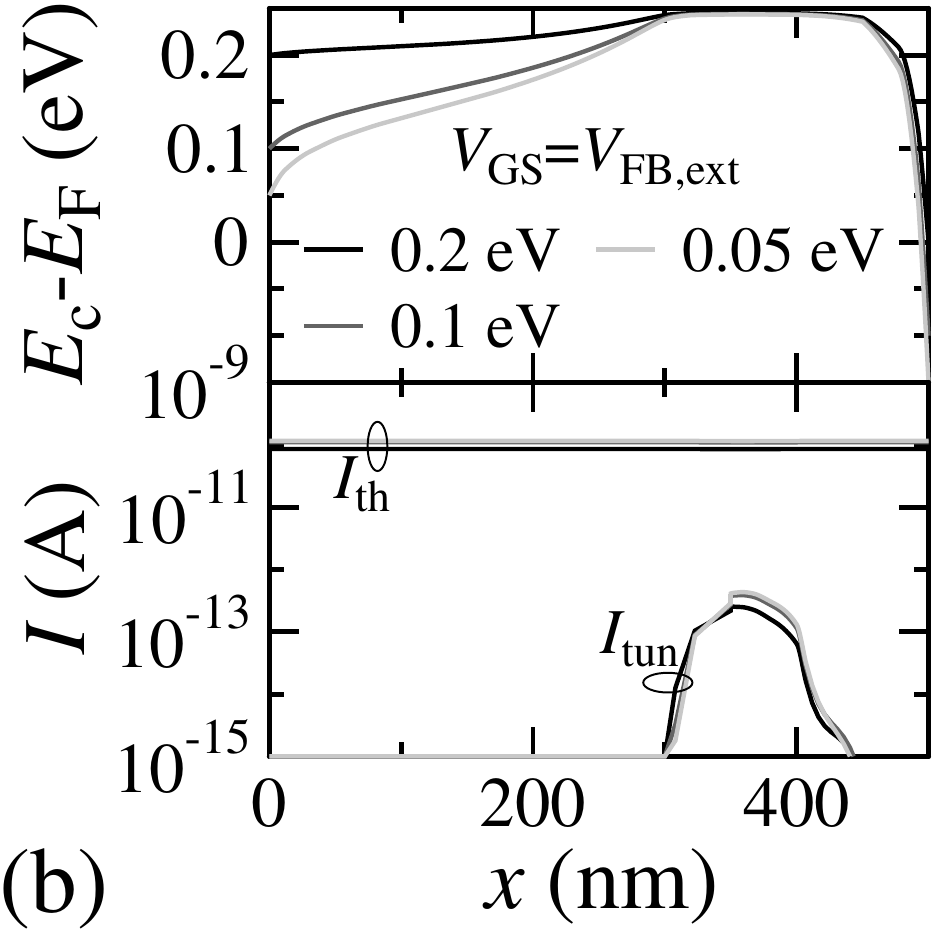}
\caption{Conduction band diagrams (top) and thermionic and tunneling currents along the device channel (bottom) of the (a) symmetric and (b) asymmetric CNTFETs. Each channel (CNT) is identified by the $\Phi_{\rm SB}$ at the metal-CNT interface set in the simulation.}
\label{fig:coos_results2}
\end{figure}

The conduction band diagrams of the individual CNT channels of both devices in Fig. \ref{fig:coos_results2} at $V_{\rm GS}=V_{\rm FB,ext}$ (identified in Fig. \ref{fig:coos_results}(b)) reveal the better electrostatic control within the the gated channel region in contrast to the spacers regions, regardless the CNT channel under study. Flat-band conditions are met at $V_{\rm FB,ext}$ only for the metal-CNT interface with highest $\Phi_{\rm SB}$ set in the simulation of the symmetric device (Fig. \ref{fig:coos_results2}(a)), whereas the onset of tunneling current indicated by $V_{\rm FB,ext}$ for the asymmetric device occurs before flat-band conditions are obtained regardless the CNT channel (Fig. \ref{fig:coos_results2}(b)). However, the lower $I_{\rm tun}$ obtained with the displaced gate device in comparison to the symmetric structure (see bottom of Fig. \ref{fig:coos_results2}) at $V_{\rm FB,ext}$ suggests that $\Phi_{\rm SB,eff}$ obtained with the former CNTFET is closer to a barrier height where pure thermionic current occurs. Hence, similar to the single-1D-channel devices \cite{PacCla17}, a test structure wtih a displaced gate hindering tunneling injection is suggested towards extracting a Schottky barrier height value closest to the true potential barrier height at metal-CNT interfaces in multi-1D-channel devices. Notice that fabricated asymmetric BG MT CNTFETs suitable for improving high-frequency performance such as the ones demonstrated elsewhere \cite{HarHer21} can be exploited for this extraction methodology as well.

\subsection{Impact of Schottky points}

\begin{figure}[!htb]
\centering
\includegraphics[height=0.1685\textheight]{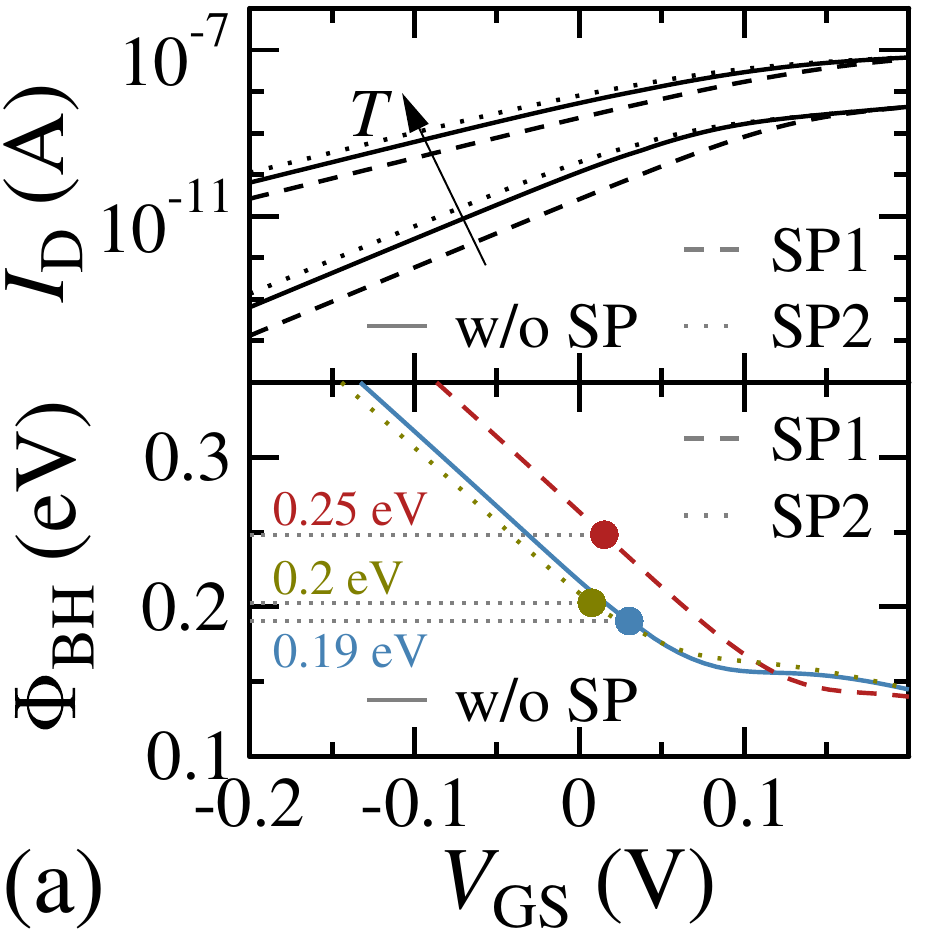}
\includegraphics[height=0.1685\textheight]{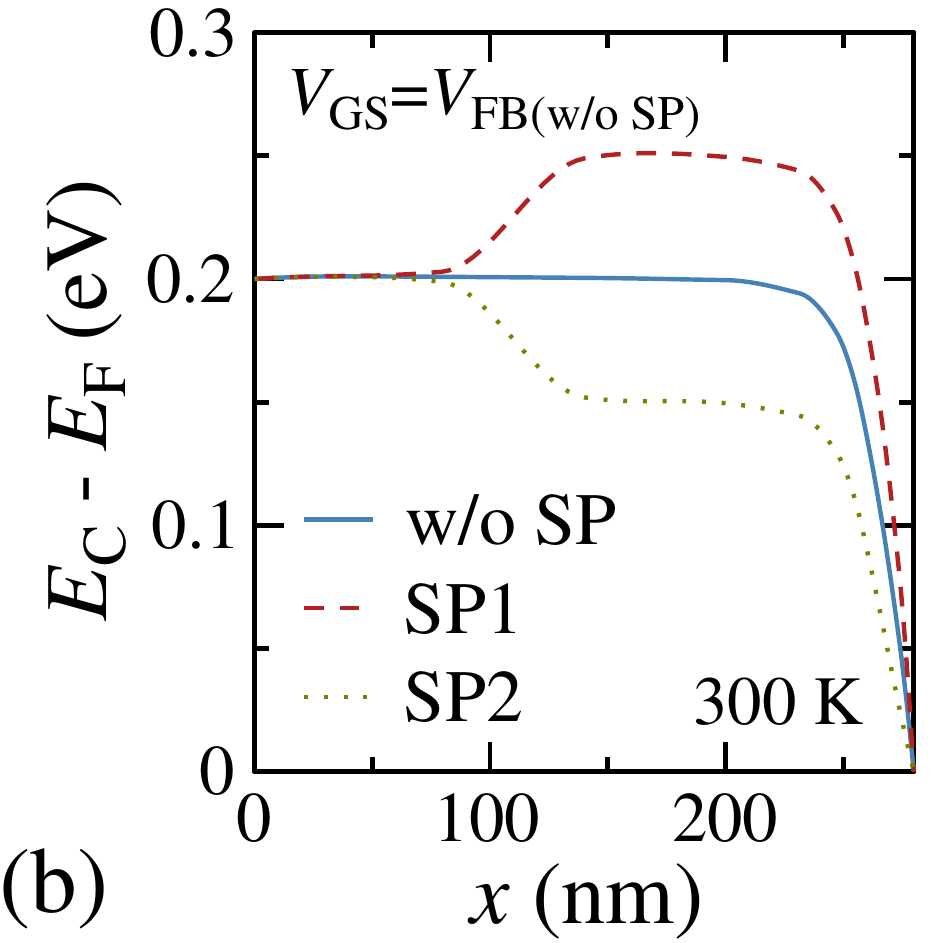}
\caption{Simulation results of BG CNTFETs with different SPs. (a) Top: transfer characteristics at \SIlist{300;500}{\kelvin} at $V_{\rm DS}=\SI{0.2}{\volt}$; bottom: potential barrier height versus $V_{\rm GS}$ obtained with 1D LBM. (b) Conduction bands at the $V_{\rm FB}$ of the devioce without SP.}
\label{fig:sp}
\end{figure}

In the present state of the technologies, CNT-based devices might suffer from crossings within the channel in non-parallel arrays in contrast to NWFETs where an improved control during fabrication enables parallel arrays. Schottky points (SPs) due to these crossings are related to a potential step in the channel electronic bands and hence, they might impact the transport\cite{BroBer19,ZorZaum21}, as well as the pure thermionic energy level required to identify $\Phi_{\rm SB}$ with 1D LBM. In this work, the impact of Schottky points on the extraction method is analyzed by means of the in-house device simulator\cite{MotSch18,ClaMot14,Mot19} previously described. Three symmetric BG CNTFETs (cf. Fig. \ref{fig:sim}(a)) with $L_{\rm ch}$ of \SI{280}{\nano\meter}, an $L_{\rm g}$ of \SI{230}{\nano\meter} and with a $\Phi_{\rm SB}$ of \SI{0.2}{\electronvolt} have been simulated with different types of SPs. Other simulation parameters are the same as the previous study (cf. Section III.B). In order to ease the discussion, a single-tube is used without loss of generality since similar tube density as in the previous simulation study (cf. Section III.B) has been considered. SPs are induced by doping a certain region of the tube, i.e., transport occurs through non-homogeneous bands. An SP1 (SP2) device has been doped towards increasing (decreasing) the energy level in a 1D channel region. A third device without SP has been simulated as well for reference purposes. Results are shown in Fig. \ref{fig:sp}. 

As shown in Fig. \ref{fig:sp}(a), the extracted $\Phi_{\rm SB}$ values (with 1D LBM) for the SP2 device and the device without SP are similar to the value of \SI{0.2}{\electronvolt} set in the simulation, whereas for the case of the SP1 device a higher value is obtained. The latter can be explained by the higher energy required not only to overcome the potential barrier at the metal-channel interface but also the potential step within the channel as shown in Fig. \ref{fig:sp} for these device in contrast to the others. Therefore, if tube crossings are present within a device, the extracted $\Phi_{\rm SB,eff}$ with 1D LBM is associated to the thermal energy required to the carriers to overcome the highest of both potential steps: at the metal-channel or at any SP within the channel. Hence, the extraction method yields the higher of the potential barriers within the channel and hence, it should be considered as a maximum limit for devices with possible tube crossings.

\section{Conclusion}

The metal-channel interfaces of fabricated FETs with arrays of CNTs and NWs with non-negligible potential barriers have been characterized here by 1D-LBM, an extraction method considering the 1D transport physics. The method extracts an effective Schottky barrier height associated to a potential separating thermionic and tunneling injection. This method overcomes the challenges encountered by conventional methods to characterize devices with non-homogeneous metal-channel interfaces. A high-performance device indicator has been also extracted such as the gate coupling coefficient. The latter helps to identify and quantify the gate control over the channel. Numerical device simulations show the improved accuracy of the characterization if a test structure is used, namely a displaced gated device. The method extracts the highest of the potential barriers in non-homogeneous electron bands within a multi-1D-channel, e.g., due to tube crossings. The methodology presented here is aim to improve the characterization and modeling of multi-1D-channel transistors where a potential barrier at the metal-channel interface can not be neglected.

%

\section*{Acknowledgements}

This work has received funding from the European Union’s Horizon 2020 research and innovation programme under grant agreements No GrapheneCore2 785219 and No GrapheneCore3 881603, from Ministerio de Ciencia, Innovación y Universidades under grant agreements RTI2018-097876-B-C21(MCIU/AEI/FEDER, UE) and FJC2020-046213-I. This  article  has been partially  funded  by  the  European Union Regional Development Fund within the framework of the ERDF Operational Program of Catalonia 2014-2020 with the support of the Department de Recerca i Universitat, with a grant of 50\% of total cost eligible. GraphCAT project reference: 001-P-001702.

The authors would like to thank Martin Friedl and Prof. Anna Fontcuberta i Morral from École Polytechnique Fédérale de Lausanne, Switzerland, for providing the experimental data of the NWFET$_{\rm EPFL}$.

\section*{Data availability statement}

The data that support the findings of this study are available from the corresponding author upon reasonable request.

\section*{Appendix. Diferrences of $\Phi_{\rm SB}$ extracted with 3D AEM and 1D LBM.}

The conceptual and mathematical differences between the novel 1D LBM and the conventional AEM for the characterization of metal-channel interfaces in 1D-devices are given here. For simplification purposes a single-1D-channel case is analyzed, however, this analysis can be extended to multiple-1D-channel devices by following the same considerations provided in Section II.

The mathematical background for AEM is based on the Richardson equation. By considering the thermionic emission of carriers (electrons) overcoming a potential barrier in 1D and an ideality factor of $\sim$\SI{1}{}, the drain current at the subthreshold region is approximately simplified to \cite{Pac19}
\setcounter{equation}{0}
\renewcommand{\theequation}{\Alph{equation}.1}
\begin{equation}
I_{\rm D} \approx A A^* T^2 \exp\left[\frac{1}{V_{\rm t}}\left(-\Phi_{\rm BH}+V_{\rm DS}\right)\right],
\end{equation}

\noindent where $A$ is an effective 3D contact area and $A^*$ is the 3D Richardson constant. By following a similar procedure as the one used to obtain Eq. (\ref{eq:phi_eff}), the expression for the potential barrier within the framework of 3D-AEM is given by

\setcounter{equation}{0}
\renewcommand{\theequation}{\Alph{equation}.2}
\begin{equation}
\Phi_{\rm BH} \approx -\frac{k_{\rm B}}{q} \frac{\partial \left[\ln \left(\frac{I_{\rm D}}{T^2}\right)\right]}{\partial T^{-1}} + V_{\rm DS}\equiv -\frac{k_{\rm B}}{q} \alpha_{\rm AEM} + V_{\rm DS},
\label{eq:phiBH_aem}
\end{equation}

\noindent which can be used to extract a $\Phi_{\rm SB}$ value. The underestimation of $\Phi_{\rm SB}$ with 3D AEM in contrast to 1D LBM can be explained by comparing Eq. (\ref{eq:phi_eff}), adapted for single channel devices, and Eq. (\ref{eq:phiBH_aem}). These differences, pointed out in Section II, are the dimension-associated exponential of the temperature in $\alpha$ and $\alpha_{\rm AEM}$ and the lack of both coupling coefficients and a $V_{\rm GS}$-associated term in the AEM case. The latter are missing issues also in an adapted 1D AEM \cite{SveSou09}. A quantitative comparison between extracted values with 1D LBM and AEM has been provided in Fig. \ref{fig:phi} as well as in previous studies\cite{PacCla17,PacRam20}

\end{document}